**Title: The first steps of adaptation of *Escherichia coli* to the gut are dominated by soft sweeps**

**Authors**:

João Barroso-Batista[1+], Ana Sousa[1+], Marta Lourenço, Marie-Louise Bergman[1], Jocelyne Demengeot[1], Karina B. Xavier[1,2] and Isabel Gordo[1*]

**Affiliations:**

[1]Instituto Gulbenkian de Ciência, Rua da Quinta Grande, 6

[2]Instituto de Tecnologia Química e Biológica, Universidade Nova de Lisboa

+equal contribution

*author for correspondence: igordo@igc.gulbenkian.pt

**Abstract:**

The accumulation of adaptive mutations is essential for survival in novel environments. However, in clonal populations with a high mutational supply, the power of natural selection is expected to be limited. This is due to clonal interference - the competition of clones carrying different beneficial mutations - which leads to the loss of many small effect mutations and fixation of large effect ones. If interference is abundant, then mechanisms for horizontal transfer of genes, which allow the immediate combination of beneficial alleles in a single background, are expected to evolve. However, the relevance of interference in natural complex environments, such as the gut, is poorly known. To address this issue, we studied the invasion of beneficial mutations responsible for *Escherichia coli*'s adaptation to the mouse gut and demonstrate the pervasiveness of clonal interference. The observed dynamics of change in frequency of beneficial mutations are consistent with soft sweeps, where a similar adaptive mutation arises repeatedly on different haplotypes without reaching fixation. The genetic basis of the adaptive mutations revealed a striking parallelism in independently evolving populations. This was mainly characterized by the insertion of transposable elements in both coding and regulatory regions of a few genes. Interestingly in most populations, we observed a complete phenotypic sweep without loss of genetic variation. The intense clonal interference during adaptation to the gut environment, here demonstrated, may be important for our understanding of the levels of strain diversity of *E. coli* inhabiting the human gut microbiota and of its recombination rate.

**Author summary:**

Adaptation to novel environments involves the accumulation of beneficial mutations. If these are rare the process will proceed slowly with each one sweeping to fixation on its own. On the contrary if they are common in clonal populations, individuals carrying different beneficial alleles will experience intense competition and only those clones carrying the stronger effect mutations will leave a future line of descent. This phenomenon is known as clonal interference and the extent to which it occurs in natural environments is unknown. One of the most complex natural environments for *E. coli* is the mammalian intestine, where it evolves in the presence of many species comprising the gut microbiota. We have studied the dynamics of adaptation of *E. coli* populations evolving in this relevant ecosystem. We show that clonal interference is pervasive in the mouse gut and that the targets of natural selection are similar in independently *E. coli* evolving populations. These results illustrate how experimental evolution in natural environments allows us to dissect the mechanisms underlying adaptation and its complex dynamics and further reveal the importance of mobile genetic elements in contributing to the adaptive diversification of bacterial populations in the gut.

**Introduction:**

Mutation is the fuel of evolution and beneficial mutations the driver of organismal adaptation. When a small group of organisms founds a new niche it will rely on *de novo* mutations to adjust to such novel environment. If the rate of emergence of new beneficial mutations is low, adaptation will proceed through the asynchronous accumulation of such mutations, one at a time. This will result in short-term polymorphism, during which the frequency of the beneficial mutation will rapidly change. Such strong selective sweeps will purge linked neutral variation, a phenomenon long recognized as the signature of selective sweeps in bacterial populations [1]. So, for reasonably small populations and for low mutation rates, population mean fitness will increase in steps, where at each step neutral variation becomes completely depleted. The adaptive walk will therefore proceed by discrete movements along the fitness landscape.

Over the years studies of microbial adaptation in different environments have repeatedly detected deviations from this simple pattern [2–6]. It is now much more commonly accepted that, in reasonably large microbial populations, many distinct adaptive mutations may arise and compete for fixation. The pattern of microbial adaptation has therefore been supportive of an evolutionary mechanism described in the sixties - the Hill-Roberston effect [7]. This has been coined clonal interference (CI) in the context of clonal bacterial populations (reviewed in [8]). This effect is theoretically expected to limit the speed of adaptation in asexual versus sexual populations [9]. A great number of beneficial mutations are lost due to interference and the winning mutations expected to have large fitness effects [8,10].

But how strong is clonal interference and what consequences does it entail? While it has been shown to occur in bacteria [5,11] and eukaryotes [3,12] in laboratory settings, its relevance in natural environments is poorly known. Interestingly, CI has recently been inferred to be an important determinant of the evolution of the influenza virus [13], suggesting that it may be pervasive in nature. The phenomenon is expected to impact the dynamics of adaptation if the effective population size ($N_e$) and/or the rate of occurrence of beneficial mutations ($U_b$) is large. More specifically, when the number of competing mutations is bigger than one, that is if $2 N_e\, U_b\, \text{Ln}(N_e\, s_b/2) > 1$ (where $s_b$ is the mean effect of a beneficial mutation [8]). Desai and Fisher [14] made the case that in extremely large populations with very high mutational inputs, a particular evolutionary regime would emerge, which they called the multiple mutation regime. In this regime haplotypes with multiple beneficial mutations are expected to arise and increase in frequency. Since the parameter values important to determine the importance and level of interference are expected to be dependent on the environment, the relevance of CI for bacterial evolution in natural conditions remains to be demonstrated.

*E. coli* K-12 has been an important model organism in many fields of biology, since its isolation from human faeces in 1922 [15]. Accumulation of mutations during its adaptation to the gut has been observed [16–18], but the fitness landscape characteristic of this environment as well as the extent of interference *E. coli* experiences is not known.

We have studied the process of accumulation of beneficial mutations and the strength of their effects in a natural environment of *E. coli*, the mouse gut. We found that CI is pervasive *in vivo* and described the genetic basis of the initial steps of adaptation, which were observed to exhibit a striking parallelism. Remarkably we have found that in the same population distinct mutations, with equivalent functional effects (that is, targeting the same gene or operon) reach detectable frequencies simultaneously. However, most get extinct after a few generations, a phenomenon called soft sweeps [19]. This translates into the occurrence of a phenotypic hard sweep without loss of variation at the genetic level.

**Results and Discussion**

**Dynamics of adaptation of *E. coli* through changes in the frequency of a neutral marker**

A genetically homogeneous population of *E. coli*, except for a chromosomally encoded fluorescence marker (see Material and Methods), was used to trace the occurrence of different adaptive mutations [2] and determine the strength of their fitness effects. This population, composed of equal amounts of two subpopulations expressing either a yellow (YFP) or cyan (CFP) fluorescent protein, was used to inoculate inbred mice orally. Subsequently, we followed the frequency of the fluorescent markers from daily collected fecal samples, for 24 days. The experiment was repeated three times to a total of 15 mice housed individually, where *E. coli* adaptation was followed.

The adaptive dynamics observed in 15 independently evolving populations are shown in Figure 1A. Some of the dynamics observed, exhibit a typical signature of CI. An initial increase in frequency of one neutral marker, followed by replacement by the subpopulation bearing another, e.g. population 1.9 (Figure 1A) was diagnostic of CI. Presumably, the markers hitchhike with successions of beneficial mutations [2]. We also observed cases in which the frequencies of the markers remained stable, (Figure 1A, populations 1.6 and 1.8), for up to 24 days. This time corresponds to approximately 400 generations, if we take the 80 minute generation time estimated in the gut [20]. Such dynamics are expected under two completely opposing scenarios: neutral evolution due to the lack of occurrence of adaptive mutations, or very intense CI caused by strong mutations of similar effect occurring in the different fluorescence backgrounds simultaneously. To distinguish between these scenarios and to further show that strong adaptive mutations are occurring, we performed *in vivo* competitive fitness assays of several clones with the ancestral strain. We tested clones from populations where a clear signal of adaptation was detected (populations 1.12 and 1.13 in Figure 1A) or was not (1.6 and 1.8 in Figure 1A). All clones tested were found to carry beneficial mutations (Figure 1B). As expected, the clones isolated after a clear sweep (Figure 1B, populations 1.12 and 1.13), showed a fitness increase. Importantly, clones sampled from the populations in which only slight changes in marker frequency were detected also showed a fitness increase (Figure 1B, populations 1.6 and 1.8). This demonstrates that the evolution process involved multiple adaptive mutations competing for fixation. Consistent with the small changes in marker frequency observed, the competitive fitnesses were similar for clones isolated from the same evolving population, but bearing distinct neutral markers (Mann-Whitney test P=0.7 population 1.6, P=0.4 population 1.8).

**The distribution of effects of mutations that contributed to adaptation**

From changes in dynamics in neutral markers one can try to estimate the important evolutionary parameters of the adaptive process [2,21–23]. One method that has been used [21] summarizes the dynamics of neutral markers in replicate experiments in two statistics**:** i) the time it takes for one of the markers to first change significantly in frequency and ii) the slope associated with that frequency change. The time for the first marker deviation corresponds to the time where the natural logarithm of the marker ratio significantly changes from the initial ratio. The slope associated with a frequency change is the slope of the linear regression of the natural logarithm of the ratio between the two markers after a significant frequency change is detected. These summary statistics lead to estimation of two effective evolutionary parameters ($U_e$ and $s_e$) [2]. Under intense clonal interference $U_e$ (the effective genomic mutation rate) will be an underestimate of $U_b$, and $s_e$ (the effective selection coefficient of beneficial mutations) an overestimate of $s_b$ [23]. For the dynamics in Figure 1, the summary statistics imply $U_e$ ~$7 \times 10^{-7}$ and $s_e$ ~0.075.

We also have used a new recently developed method [22] to estimate the fitness of the haplotypes that segregate at sufficiently high frequency to lead to changes in marker dynamics. This method takes the frequencies of the neutral markers across all the time points of the experiment and allows for the occurrence of CI. We started by estimating the distribution of fitness effects of the minimum number of haplotypes, as proposed by Illingworth and Mustonen [22], to fit the marker frequency data. Assuming the simplest possible model of Darwinian selection, with constant selection across space and time (admittedly an oversimplified view of the gut), we fitted a series of models, which differ in the number of beneficial mutations. Starting from the simplest minimal model where only one beneficial mutation occurs, we sequentially increase the number of mutations until a maximum of five. For each model this approach fits by maximum likelihood the observed frequency changes at a marker locus with two alleles (our fluorescent alleles) and then chooses the simplest possible model. The predictions described faithfully the empirical data in terms of marker frequencies (lines in Figure 1A). With this method the distribution of haplotype fitnesses across all populations, can be determined (Figure 2). The inferred fitness effects of segregating haplotypes are quite large with a mean of (15%) and some haplotypes leading to increases in fitness of more than 30%. We should note that given the intense CI observed, this approach may miss some of the mutations contributing to the dynamics (see below).

**The genetic basis of the initial adaptations to the mouse gut**

To characterize genetically the first steps of the adaptive process *in vivo*, we performed whole genome sequencing of the ancestral and independently evolved clones. Each evolved clone was isolated from a different mouse after colonization for 24 days (~400 generations). The mean number of mutations per evolved clone was 2.3, with a minimum of 1 and a maximum of 4 (Figure 3 and Table S2, see also Table S1).

The genetic analysis showed that similar and parallel adaptive paths were taken during the initial adaptation to the gut (Figure 3). The most striking recurrent event was detected at the level of mutations in the *gat* operon, which occurred in all the sequenced clones (79% IS insertions and 21% small indels). The *gat* operon consists of six genes that collectively allow for galactitol metabolism [24]. We found that galactitol had an inhibitory effect on the ancestral strain when it grows in minimal medium with glycerol (Figure S1). However in all the evolved clones (carrying mutations either in the coding region of *gatA*, *gatC*, *gatZ* or in the regulatory region of *gatY* (see Figure 3)) this inhibition was surpassed. All these clones exhibited a *gat*-negative phenotype, inability to metabolize galactitol. Since galactitol is part of the host's metabolism of galactose, *E. coli* might be frequently exposed to this compound in the gut. This may have selected for the *gat*-negative phenotype.

Parallelism at the gene level was also observed: four clones carried mutations in *srlR*, which is a DNA-binding transcription factor that represses the *srl* operon involved in sorbitol metabolism [25]. At least two of the mutations inactivated the gene, since these caused a frameshift and a stop codon (Table S2). The inactivation of *srlR* leads to the constitutive expression of the *srl* operon [25], thus these mutants are expected to readily consume sorbitol, even at low concentrations. This might represent an advantage when facing limiting nutrients and strong competition, compatible with the hypothesis that the ability to use several limiting carbon sources is an important advantage for *E. coli* inhabiting the mammalian gut [26,27].

Two other parallelisms in the regulatory regions of *dcuB* (three clones) and *focA* (two clones) were observed. Both of these genes encode transmembrane transporters whose expression occurs almost exclusively in anaerobic conditions. The Dcu system mediates the uptake, exchange and efflux of $C_4$-dicarboxylic acids [28]; fumarate is a C4-dicarboxylate acid and is the most important anaerobic electron acceptor for *E. coli* in the intestine [29]. Additionally, *focA* is a putative transporter of formate [30] and is the "signature compound" of *E. coli* anaerobic metabolism [31]. During anaerobic growth, pyruvate is cleaved into formate that is subsequently used as a major electron donor for the anaerobic reduction of fumarate [32] and nitrate or nitrite [33]. Hence, given their roles in anaerobic respiration it would be expected that *dcuB* and *focA* are under selective pressure in the gut and that the mutations observed may be important for regulating anaerobic respiration of *E. coli* in the gut.

Interestingly, one clone carried a non-synonymous mutation in *arcA*, a global regulator that governs respiratory flexibility of *E. coli*. *arcA* allows controlled switching between aerobic and anaerobic respiratory genes during fluctuating oxygen conditions, a trait crucial for *E. coli* to efficiently compete with the gut flora composed essentially of anaerobes [29,34].

Previous studies have found increased expression of *dcu*, *gat* and *srl* regulons during growth of *E. coli* on mucus [16]. Together with our findings of mutations on these genes, suggests that these regulons are under strong selection.

One common adaptation in *E. coli* reported to occur in the mouse gut is decreased motility [17,27,35], which typically occurs in strains that are hypermotile as a result of an IS element upstream of the *flhDC* operon. Since our strain is not hypermotile, it is not surprising that mutations in this region were not observed in our study.

Taken together, our genetic analyses revealed that the initial adaptive steps of *E. coli* to the mouse gut involved gene inactivation or modulation through IS elements. In terms of the relative contribution of regulatory versus coding regions to adaptation [36,37], we found that half of the IS insertions occurred in regulatory regions and one third of all mutations were located in these regions. We therefore observe that both changes in regulatory and coding regions contributed to *E. coli* adaptation to the gut. We also note that the first step of adaptation involved loss of function mutations (namely in the *gat* operon), which supports the recent view that null mutations may play an important role in early adaptation to new environments [38].

Previous studies [4,39] focusing on evolution of *E. coli* populations in glucose-limited chemostats, a device used to simulate the human gut [40], during a similar period of time (26 days), found a high degree of parallelism just as we observe here. Furthermore extensive phenotypic diversity was observed [39] and whole genome sequencing identified similar numbers of substitutions and a contribution of IS elements to the adaptive process [41]. However the targets of selection in glucose-limited chemostats were very different from those found during its adaptation to the gut, reflecting the distinct specific selective pressures *E. coli* is subjected to in the different environments.

**Soft sweeps at the *gat* operon and the multiple mutation regime of adaptation**

Given the striking parallelism that occurred in the *gat* operon, through knockout mutations, we sought to determine the timing of its emergence in each of the populations. We followed the dynamics of the *gat*-negative phenotype during adaptation (Figure 4). In eight populations the advantageous mutation could be detected just 2 days post-colonization. In one population an extraordinarily rapid phenotypic sweep could be seen: a population where initially the majority of clones were *gat*-positive changed completely its phenotype within 2 days (Figure 4, population 1.5), showing the strong benefit associated with this phenotype.

In most populations the increase in frequency of the *gat*-negative phenotype was followed by a change in marker frequency (Figure S2A), consistent with the expected dynamics of a strong beneficial mutation occurring in a given fluorescent background sweeping to fixation. However in some populations fixation of the *gat*-negative phenotype was observed without fixation of the neutral marker (Figure S2B). This suggests that independent mutations causing the same phenotype have occurred, *i.e.* clonal competition emerging from multiple alleles at the same locus with similar selective effects is driving these dynamics.

To gain further insight into the regime of interference detected by the change in frequency of the neutral markers, we studied polymorphism levels of four of the adapting populations. We followed the change in frequency of haplotypes at different points in time in two populations that showed a rapid change in marker frequency (populations 1.5 and 1.12) and in two other populations where the neutral marker was kept polymorphic throughout the period studied (1.1 and 1.11) (Tables S3 to S6). Each haplotype was the combined genotype at 7 selected loci: *gatY, gatZ, gatA, gatC, srlR, dcuB* and *focA*. These loci were previously observed to be mutational targets in two or more of the sequenced independently evolved clones (see Fig. 3), which we interpret as important players in adaptation to the gut, and thus likely exhibiting segregating polymorphisms in most populations. We note that we determined the haplotype structure of a sample of clones (between 20 and 40 clones per time point, per population) based on the mutations found in the sequenced clones, that is, we looked for IS insertions in *gatY, gatZ, gatA, gatC, dcuB* and *focA*, and SNPs or small indels in *gatZ, gatC and srlR* (see Methods). The presence of an extra mutation, a duplication of ~150Kb, was also tested in population 1.12.

Figure 5 shows the dynamics of these haplotypes in the populations studied. Extensive haplotype diversity is found in all populations but to a lesser extent in populations 1.12 and 1.5. For example, in the first ~100 generations we observe 16 and 8 haplotypes segregating in populations 1.1 and 1.11, but only 3 and 5 in populations 1.12 and 1.5, respectively. This observation is concordant with the neutral marker dynamics, where populations 1.12 and 1.5 showed a rapid sweep of one of the markers while the other two (1.1 and 1.11) maintained the ratio of these markers closer to 50% for a longer period. The first mutation detected, without exception, impairs galactitol metabolism, thus surpassing the vital inhibitory effect of this compound on *E coli*'s growth. We observed multiple soft sweeps of mutations, with possibly similar fitness benefit, occurring independently in the *gat* operon and competing for fixation. Over time a second adaptive mutation occurred and promoted the increase in the frequency of haplotypes carrying more than one beneficial mutation. Overall, the adaptive process appears non-mutation limiting, allowing polymorphism levels within populations to be kept high until the end of the period studied. A closer look to the haplotype dynamics points to the strong possibility of further mutations occurring beyond the ones we have targeted. For example, at generation 126 in population 1.12, the increase in frequency of the neutral allele *yfp* seems to be caused by some other mutations besides *gatC*. Furthermore the occurrence of a large duplication in the same population appears to cause a beneficial effect. Overall we found ~6% of triple mutants. These were all *gat*-negative and carried either a *srlR and dcuB* mutation or a *srlR and focA* mutation. *dcuB* and *focA* were never found in combination, which can be possibly explained by their common involvement in anaerobic respiration modulation, thus fulfilling related functions.

We next sought to estimate the strength of selection associated with new beneficial alleles that emerge and escape stochastic loss from the initial change in frequency of the *gat*-negative phenotype. If we take population 1.12, where only one allele was detected in the *gat* operon, we can estimate the selective advantage conferred by this mutation ($s_{gat}$) from the slope of the Ln($x$/(1-$x$)) along time, where $x$ is the frequency of the *gat*-negative phenotype. When doing so we find that $s_{gat}$ = 0.07 (±0.01) (Figure 4 inset). In the other populations the strength of selection is more difficult to estimate due to the emergence of many distinct alleles, which leads to an underestimate of their beneficial effect due to CI.

**Test for negative frequency-dependent selection**

Negative frequency dependent selection, *i.e.*, fitness advantage of clones when at low frequency and fitness disadvantage when at high frequency, is known to lead to maintenance of genetic diversity [11,39,42]. Since the gut is a complex environment [43] where non-transitive ecological interactions may occur, we tested for variation in fitness effects with initial frequency of evolved clones. Negative frequency-dependent interactions could explain the long-term maintenance at low frequency of some of the fluorescent subpopulations observed in the adaptive dynamics (Figure 1A, populations 1.5, 1.12 and 1.13). We tested population 1.13 for this type of selection, where CFP marker increased till 0.95 but never achieved fixation. In Figure S3 we show the results of the tested adapted clones in competitive fitness assays at different initial frequencies. While we observed that clones have an advantage when rare we also observe that such selective advantage is present when at high frequency. No disadvantage was found at any of the initial frequencies tested, which suggests that negative frequency dependent selection is not a major force operating in this population.

**No evidence for emergence of mutators**

Our finding that mutations of large beneficial effects can accumulate rapidly *in vivo* suggested to us that mutators could have emerged. Mutators may play an important role in bacterial evolvability and their emergence typically involves mutations in genes of the DNA repair system [44]. Depending on the gene mutated a mutator bacteria can acquire an increase in its genome-wide mutation rate of 10 up to 10 000 fold [45]. Indeed co-colonization of wild-type and mutator bacteria in germ free mice [46] has shown that mutators can increase in frequency by hitchhiking with beneficial mutations to which they are genetically linked to. Furthermore mutators have been observed to sometimes emerge *de novo* in the early stages of adaptation to glucose-limited environments in laboratory experiments [47,48]. We therefore tested for an increased rate of mutation through fluctuation assays of several adapted clones. These tests however provided no evidence for significant increases in genomic mutation rate (Figure S4).

In conclusion, we demonstrate here that *E. coli* MG1655 adapts very rapidly to the intestine of streptomycin-treated mice. The first steps of adaptation of *E. coli* to the mouse gut are dominated by soft sweeps and adaptive mutations of large effect. We observed a regime of intense clonal interference where haplotypes carrying more than one beneficial mutation compete for fixation. This is the regime assumed under the Fisher-Muller theory that predicts that recombination speeds up adaptation by reducing competition between beneficial mutations. This data therefore provides support for the Fisher–Muller effect as an important mechanism for the maintenance of homologous gene recombination in bacteria [49]. It shows that *E. coli* can adapt to this complex ecosystem very fast: within two days a phenotypic replacement could be detected. The first beneficial mutations targeted the same gene or operon and transposable elements made a substantial contribution to adaptation. These results demonstrate the remarkable adaptive potential of bacteria in one of the most complex environments of their natural ecology and may have important consequences for our understanding of the species and strain diversity in the gut microbiota.

**Material and Methods:**

**Ethics statement**

All experiments involving animals were approved by the Institutional Ethics Committee at the Instituto Gulbenkian de Ciência (project nr. A009/2010 with approval date 2010/10/15), following the Portuguese legislation (PORT 1005/92), which complies with the European Directive 86/609/EEC of the European Council.

### *E. coli* Strains

All strains used were derived from *Escherichia coli* K-12, strain MG1655 [44].

Strains DM08-YFP and DM09-CFP (MG1655, *galK::YFP/CFP* amp$^R$ *(pZ12)*, str$^R$ *(rpsl150), ΔlacIZYA*) were used in the initial colonization experiment. These strains contain the yellow (*yfp*) or cyan (*cfp*) fluorescent genes linked to amp$^R$ in the *galK* locus under the control of a *lac* promoter and were obtained by P1 transduction from previously constructed strains [2]. To ensure constitutive expression of the fluorescent proteins the *lac* operon was deleted using the Datsenko and Wanner method [50].

### Fluorescent marker dynamics during mouse colonization

To study *E. coli* adaptation to the gut we used a streptomycin-treated mouse colonization model [43]. Briefly, 6- to 8-week old C57BL/6 male mice raised in specific pathogen free (SPF) conditions were given autoclaved drinking water containing streptomycin (5g/L) for one day. After 4 hours of starvation for water and food the animals were gavaged with 100 µl of a suspension of $10^8$ colony forming units (CFUs) of a mixture of YFP- and CFP-labeled bacteria (ratio 1:1) grown at 37°C in brain heart infusion medium to $OD_{600}$ of 2. After gavage, the animals were housed separately and both food and water containing streptomycin were returned to them. Mice fecal pellets were collected for 24 days, diluted in PBS and plated in Luria Broth agar (LB agar). Plates were incubated overnight and the frequencies of CFP- or YFP-labeled bacteria were assessed by counting the fluorescent colonies with the help of a fluorescent stereoscope (SteREO Lumar, Carl Zeiss). A sample of each collected fecal pellet was daily stored in 15% glycerol at -80ºC for future experiments.

For the initial colonization a group of 5 mice was gavaged and followed for 24 days, and this procedure was repeated for two more groups of 5 mice. Thus a total of 15 mice were analyzed (Figure 1A).

**Competitive fitness assays *in vivo* to test for adaptation**

We measured the relative fitness *in vivo* by bacterial clones isolated from mouse fecal samples after 24 days of colonization (approximately 400 generations). Individual colonies or mixtures of approximately 30 colonies with the same fluorescent marker (population of clones) were picked from mouse fecal platings, grown in 10 ml of Luria Broth medium (LB) supplemented with ampicillin (100 µg/ml) and streptomycin (100 µg/ml) and stored in 15% glycerol at -80ºC.

The isolated clones were competed against the ancestral strain labeled with the opposite fluorescent marker, at a ratio of 1 to 1, in 1-2 day co-colonization experiments following the same procedure described for the evolution experiments. The selective coefficient (fitness gain) of these clones *in vivo* (presented in Figure 1B) was estimated as:

$$s_b = \ln\left(\frac{Rf_{ev/anc}}{Ri_{ev/anc}}\right)/t$$

where $s_b$ is the selective coefficient of the evolved clone, $Rf_{ev/anc}$ and $Ri_{ev/anc}$ are the ratios of evolved to ancestral bacteria in the end (*f*) or in the beginning (*i*) of the competition and *t* is the number of generations per day. We assume *t*=18, in accordance with the 80 minute generation time estimated in previous studies on *E. coli* colonization of streptomycin-treated mouse [20,43,51].

**Estimation of rate of beneficial mutations**

To infer the effective evolutionary parameters ($U_e$ and $s_e$) we used the method described in [21], available online at

[http://barricklab.org/twiki/bin/view/Lab/ToolsMarkerDivergence](http://barricklab.org/twiki/bin/view/Lab/ToolsMarkerDivergence). This procedure involves three main steps: simulate families of marker trajectories at many combinations of mutation rate (*U*) and selection coefficient (*s*), summarize the simulated and experimental marker trajectories in the statistics $\tau_e$ and $\alpha_e$ (that represent the time of divergence of marker frequency and the rate of divergence, respectively); compare the distributions of $\tau_e$ and $\alpha_e$ of the simulated dynamics and the experimental data and find the values of *U* and *s* that best explain the experimental marker dynamics.

Using the model described in [21] we simulated sets of 100 replicate populations evolved under different parameter combinations of *U* and *s*. *U* ranged from $10^{-8}$ to $10^{-4}$ (with increments of $10^{-8}$, $10^{-7}$, $10^{-6}$, $10^{-5}$ and $10^{-4}$) and *s* ranged from 0.07 to 0.1 (with increments of 0.005). We assumed a population size of $10^7$, based on the weight of feces per day (1.5 ± 0.12 (2 s.e.m.) grams) that a mouse produces, on the observed number of CFUs (per gram of feces) along the experiments (which remained stable over the length of the experiment around $10^8$), and on the number of generations per day, which is 18 [20]. A population size of the same order of magnitude was measured by Leatham-Jensen *et al.* [52] in the intestinal mucus, where the majority of bacterial division occurs [43].

This simulated data consist of the logarithm of the ratio of the two marker frequencies (Rf = $f_{YFP}$(ti)/($f_{CFP}$(ti))) at several time points (ti), (ln(Rf(ti))).

We then used both the simulated and experimental marker dynamics as input in the program marker_divergence_fit.pl, that summarizes the evolutionary dynamics in two statistics: $\tau_e$ and $\alpha_e$. The first is the time, $\tau_e$, where a significant deviation of ln(Rf($\tau_e$)) from ln(Rf($t$ = 0)) occurs. The second is the rate of change of ln(Rf($t$)) with time, that is, $\tau_e$ sets the time of divergence of marker frequency and $\alpha_e$ the rate of divergence. Each replicate population is summarized by a single value of $\tau_e$ and $\alpha_e$, and the *n* replicate populations (characterized by a given combination (*U*, *S*)) result in a distribution of $T(\tau_e)$ and $A(\alpha_e)$. These distributions are then compared, using the program

marker_divergence_significance.pl, to the distributions of $τ_e$ and $α_e$ that summarize the experimental data To($τ_e$) and Ao($α_e$) using a two-dimensional Kolmogorov–Smirnov to test the fit between the simulated data and the pseudo-observed data. The combination ($U$, $S$) that gives rise to the highest $P$ value is taken as $U_e$ and $S_e$, even when the hypothesis that the distributions are different cannot be rejected.

**Estimation of distribution of haplotypes fitness**

To estimate the fitness effects of beneficial mutations that may contribute to adaptation we used a maximum likelihood approach to infer the number, fitness effect and time of establishment of beneficial mutations in an asexual population from individual marker frequencies trajectories [22]. This method, which is available on web page: http://www.sanger.ac.uk/resources/software/optimist/, makes no *a priori* assumptions about the distribution of mutation effects. It has been tested against simulated data and has been shown to be able to: reproduce complex trajectories of neutral markers, make accurate estimates of the distribution of haplotype fitness effects, and provide good estimates of the number of mutations segregating at high frequencies, at least for values of mutation rates that are not very large. We note that this method may miss many small effect beneficial mutations if these indeed occur at high rates.

It assumes an initial population of cells comprising two haplotypes with equal fitness, corresponding to two neutral markers. A beneficial mutation occurring in one of the backgrounds is represented as a new haplotype, which is assumed to exist at a given time ($Tb_i$) and with a frequency of 0.001 in the population. Since this new haplotype has increased fitness ($W_i$) relative to the initial wild-type population, its frequency will increase, leading to an increase in frequency of the fluorescent marker population in which the mutation has occurred. The simplest model that can be assumed is that where a single beneficial mutation occurs. That will imply a given time of establishment and effect of the mutation that maximize the probability of observing the marker frequency data. The second model that is considered is one where two mutations can occur, which will lead to a given likelihood of the data. Models assuming different numbers of beneficial mutations can thus be compared based on their likelihood score using Akaike information criteria [22]. With this method we obtain a distribution of haplotype fitnesses with the minimal number of mutations that can explain the marker frequency dynamics ($\omega_h$).

**Whole genome re-sequencing and mutation prediction**

After 24 days of gut colonization one clone from populations 1.1 to 1.14 and the two ancestors (MG1655-YFP and MG1655-CFP) were isolated and used to seed 10 mL of LB (Line 1.15 was not analyzed since the mouse from this line died at that time point). These cultures were then grown at 37ºC with agitation. Subsequently DNA was isolated following a previously described protocol [53]. The DNA library construction and sequencing was carried out by BGI. Each sample was pair-end sequenced on an Illumina HiSeq 2000. Standard procedures produced data sets of Illumina paired-end 90bp read pairs with insert size (including read length) of ~470bp. Genome sequencing data have been deposited in the NCBI Read Archive, http://www.ncbi.nlm.nih.gov/sra (accession no. SRA062068).

Mutations were identified using the BRESEQ pipeline [54]. To detect potential duplication events we used ssaha2 [55] with the paired-end information. This is a stringent analysis that maps reads only to their unique match (with less than 3 mismatches) on the reference genome. Sequence coverage along the genome was assessed with a 250 bp window and corrected for GC% composition by normalizing by the mean coverage of regions with the same GC%. We then looked for regions with high differences (>1.4) in coverage. Large deletions were identified based on the absence of coverage. For additional verification of mutations predicted by BRESEQ, we also used the software IGV (version 2.1) [56].

*Ancestral genome:* The sequence reads from MG1655 were mapped to the reference strain [57]. The two ancestors carried the mutations listed in Table S1. The mutations underlined were present in the YFP ancestor but not in the CFP. The sequences of the 14 sequenced clones were then interrogated against this ancestral genome and the mutations identified are listed in Table S2.

**Test for the advantage of the evolved clones in the presence of galactitol**

All clones that were sequenced carried mutations in the *gat* operon. To test for any fitness advantage of these evolved clones, we measured their growth in M9 minimal medium (MM) supplemented with glycerol and with glycerol and galactitol, and compared growth in both conditions. All growth curves were conducted in 96-well plates incubated at 37ºC with aeration (Thermoshaker PHMP-4, Grant). After a first overnight growth in MM with glycerol (0.4%) the cultures were normalized to the same $OD_{600}$ (approximately 0.1) and diluted 100-fold. We used 5 µl of the $10^{-2}$ dilution to inoculate in triplicate wells containing MM supplemented either with glycerol (0.4%) or glycerol (0.4%) and galactitol (0.4%). Optical density at 600 nm was monitored for 36 hours using a microplate reader (Victor3, PerkinElmer). The results are shown in Figure S1.

**Emergence and dynamics of mutations in the *gat* operon**

To investigate the dynamics of appearance and expansion of beneficial mutations in the *gat* operon we determined the frequency of bacteria unable to metabolize galactitol (*gat*-negative phenotype) within a given population of the evolution experiment.

For each population, a sample of the frozen fecal pellets was diluted in PBS and plated in Mac Conkey agar supplemented with 1% of galactitol and streptomycin (100 µg/ml); a differential medium used to monitor the ability of bacteria to use galactitol. Plates were incubated for 20 hours at 28ºC. The frequency of galactitol mutants for each time point was estimated by counting the number of white (galactitol mutants) and red colonies in Mac Conkey-galactitol plates.

**Identification of adaptive mutations and estimate of haplotype frequencies in populations 1.1, 1.5, 1.11 and 1.12 during 24 days of adaptation to the mouse gut**

In order to identify the adaptive mutations and estimate the haplotype frequencies, between 20 and 40 clones were analyzed per time point per population. The presence of the adaptive mutations was queried by target PCR. The increase in size of the PCR product was indicative of the presence of an IS element. The detection of SNPs was done by target Sanger sequencing using the primers listed in Table S7.

PCR reactions were performed in a total volume of 50 µl containing 1 µl bacterial culture, 10 µM of each primer (forward and reverse), 200 µM dNTPs, 1 U Taq polymerase and 1X Taq polymerase buffer. The PCR reaction conditions were as follows: 95ºC for 3 min followed by 34 cycles of 95 ºC for 30 s, 60 ºC for 30 s, 72 ºC for 2 min, followed by 72ºC for 5min. *srlR*, *gatC* and *gatZ* were sequenced directly from the PCR products using the primers listed in Table S7.

To detect an insertion of 1 bp in *gatC*, the gene was PCR amplified and then submitted to an enzymatic restriction with MvaI enzyme. The mutant was identified based on the fragment restriction profile.

The 150 Kb duplication was detected by amplification of the new junction formed by the tandem duplication. The PCR was performed using the same conditions described before and the primers listed in Table S7.

**Testing for negative frequency dependent selection**

To test for this form of selection we isolated around 30 YFP and 30 CFP clones from fecal samples after 24 days of colonization of population 1.13. We then performed competitive fitness assays *in vivo*. The mixture of the 30 YFP clones was competed against the mixture of the 30 CFP clones at three initial ratios (1:1, 10:1 and 1:100) for 2 days, using the procedure described for the evolution experiments. The results are shown in Figure S3.

**Test for increased mutation rate**

To test for the possible emergence of mutators during adaptation to the gut we determined the frequency of rifampicin-resistant mutants, in each of the evolved populations. We grew pre-cultures of the evolved populations by inoculating in 10 ml of LB a sample of each population from the last day of the experiments (approximately 400 generations of gut colonization). Pre-cultures were grown overnight at 37ºC with aeration in the presence of ampicillin (100 µg/ml) and streptomycin (100 µg/ml). The pre-cultures were then diluted and approximately 1000 cells (10 µl of a $10^{-4}$ dilution) were inoculated in triplicate in 1 ml of LB and incubated overnight. Aliquots of each tube were plated in LB agar and LB agar supplemented with rifampicin (100 µg/ml) and incubated overnight at 37ºC. The frequency of mutation to rifampicin resistance was calculated as the ratio between the number of rifampicin resistant mutants and the total number of individuals in each population.

**Acknowledgments:**

We thank J. Xavier, L. Wahl, A. Athanasiadis, A. Coutinho, J. Thompson and Gordo's lab for critically reading the manuscript and D. Sobral for help with sequence analysis.

**References:**

1. Atwood KC, Schneider LK, Ryan FJ (1951) Periodic selection in *Escherichia coli*. Proc Natl Acad Sci USA 37: 146–155.

2. Hegreness M, Shoresh N, Hartl D, Kishony R (2006) An equivalence principle for the incorporation of favorable mutations in asexual populations. Science 311: 1615–1617. doi:311/5767/1615.

3. Kao KC, Sherlock G (2008) Molecular characterization of clonal interference during adaptive evolution in asexual populations of *Saccharomyces cerevisiae*. Nature Genet 40: 1499–1504. doi:ng.280.

4. Maharjan R, Seeto S, Notley-McRobb L, Ferenci T (2006) Clonal adaptive radiation in a constant environment. Science 313: 514–517. doi:10.1126/science.1129865.

5. Perfeito L, Fernandes L, Mota C, Gordo I (2007) Adaptive mutations in bacteria: high rate and small effects. Science 317: 813–815. doi:317/5839/813.

6. Woods RJ, Barrick JE, Cooper TF, Shrestha U, Kauth MR, et al. (2011) Second-order selection for evolvability in a large *Escherichia coli* population. Science 331: 1433–1436. doi:10.1126/science.1198914.

7. Hill WG, Robertson A (1966) The effect of linkage on limits to artificial selection. Genet Res 8: 269–294.

8. Sniegowski PD, Gerrish PJ (2010) Beneficial mutations and the dynamics of adaptation in asexual populations. Philos Trans R Soc London B Biol Sci 365: 1255–1263. doi:365/1544/1255.

9. Barton N, Partridge L (2000) Limits to natural selection. Bioessays 22: 1075–1084. doi:10.1002/1521-1878(200012)22:12<1075::AID-BIES5>3.0.CO;2-M.

10. Gordo I, Perfeito L, Sousa A (2011) Fitness effects of mutations in bacteria. J Mol Microbiol Biotechnol 21: 20–35. doi:000332747.

11. Herron MD, Doebeli M (2013) Parallel evolutionary dynamics of adaptive diversification in *Escherichia coli*. PLoS Biol 11: e1001490. doi:10.1371/journal.pbio.1001490.

12. Desai MM, Fisher DS, Murray AW (2007) The speed of evolution and maintenance of variation in asexual populations. Curr Biol 17: 385–394. doi:S0960-9822(07)00984-0.

13. Strelkowa N, Lässig M (2012) Clonal interference in the evolution of influenza. Genetics 192: 671–682. doi:10.1534/genetics.112.143396.

14. Desai MM, Fisher DS (2007) Beneficial mutation selection balance and the effect of linkage on positive selection. Genetics 176: 1759–1798. doi:genetics.106.067678.

15. Lederberg J (2004) *E. coli* K-12. Microbiol Today 31: 116.

16. Fabich AJ, Leatham MP, Grissom JE, Wiley G, Lai H, et al. (2011) Genotype and phenotypes of an intestine-adapted *Escherichia coli* K-12 mutant selected by animal

passage for superior colonization. Infect Immun 79: 2430–2439. doi:10.1128/IAI.01199-10.

17. Leatham MP, Stevenson SJ, Gauger EJ, Krogfelt KA, Lins JJ, et al. (2005) Mouse intestine selects nonmotile *flhDC* mutants of *Escherichia coli* MG1655 with increased colonizing ability and better utilization of carbon sources. Infect Immun 73: 8039–8049. doi:10.1128/IAI.73.12.8039.

18. De Paepe M, Gaboriau-Routhiau V, Rainteau D, Rakotobe S, Taddei F, et al. (2011) Trade-off between bile resistance and nutritional competence drives *Escherichia coli* diversification in the mouse gut. PLoS Genet 7: e1002107. doi:10.1371/journal.pgen.1002107 PGENETICS-D-11-00191.

19. Lee MC, Marx CJ (2013) Synchronous waves of failed soft sweeps in the laboratory: remarkably rampant clonal interference of alleles at a single locus. Genetics 193: 943–952. doi:10.1534/genetics.112.148502.

20. Poulsen LK, Licht TR, Rang C, Krogfelt KA, Molin S (1995) Physiological state of *Escherichia coli* BJ4 growing in the large intestines of streptomycin-treated mice. J Bacteriol 177: 5840–5845.

21. Barrick JE, Kauth MR, Strelioff CC, Lenski RE (2010) *Escherichia coli rpoB* mutants have increased evolvability in proportion to their fitness defects. Mol Biol Evol 27: 1338–1347. doi:10.1093/molbev/msq024.

22. Illingworth CJR, Mustonen V (2012) A method to infer positive selection from marker dynamics in an asexual population. Bioinformatics 28: 831–837. doi:10.1093/bioinformatics/btr722.

23. Sousa JAM de, Campos PRA, Gordo I (2013) An ABC method for estimating the rte and distribution of effects of beneficial mutations. Genome Biol Evol 5: 794–806. doi:10.1093/gbe/evt045.

24. Nobelmann B, Lengeler JW (1996) Molecular analysis of the *gat* genes from *Escherichia coli* and of their roles in galactitol transport and metabolism. J Bacteriol 178: 6790–6795.

25. Yamada M, Saier MH Jr (1987) Physical and genetic characterization of the glucitol operon in *Escherichia coli*. J Bacteriol 169: 2990–2994.

26. Chang DE, Smalley DJ, Tucker DL, Leatham MP, Norris WE, et al. (2004) Carbon nutrition of *Escherichia coli* in the mouse intestine. Proc Natl Acad Sci USA 101: 7427–7432. doi:10.1073/pnas.0307888101.

27. Gauger EJ, Leatham MP, Mercado-Lubo R, Laux DC, Conway T, et al. (2007) Role of motility and the *flhDC* operon in *Escherichia coli* MG1655 colonization of the mouse intestine. Infect Immun 75: 3315–3324. doi:10.1128/IAI.00052-07.

28. Engel P, Krämer R, Unden G (1994) Transport of C4-dicarboxylates by anaerobically grown *Escherichia coli.* Energetics and mechanism of exchange, uptake and efflux. Eur J Biochem 222: 605–614.

29. Jones SA, Gibson T, Maltby RC, Chowdhury FZ, Stewart V, et al. (2011) Anaerobic respiration of *Escherichia coli* in the mouse intestine. Infect Immun 79: 4218–4226. doi:10.1128/IAI.05395-11.

30. Suppmann B, Sawers G (1994) Isolation and characterization of hypophosphite-resistant mutants of *Escherichia coli*: identification of the FocA protein, encoded by the *pfl* operon, as a putative formate transporter. Mol Microbiol 11: 965–982.

31. Leonhartsberger S, Korsa I, Böck A (2002) The molecular biology of formate metabolism in enterobacteria. J Mol Microbiol Biotechnol 4: 269–276.

32. Macy J, Kulla H, Gottschalk G (1976) H2-dependent anaerobic growth of *Escherichia coli* on L-malate: succinate formation. J Bacteriol 125: 423–428.

33. Wimpenny JW, Cole JA (1967) The regulation of metabolism in facultative bacteria. 3. The effect of nitrate. Biochim Biophys Acta 148: 233–242.

34. Jones SA, Chowdhury FZ, Fabich AJ, Anderson A, Schreiner DM, et al. (2007) Respiration of *Escherichia coli* in the mouse intestine. Infect Immun 75: 4891–4899. doi:10.1128/IAI.00484-07.

35. Giraud A, Arous S, De Paepe M, Gaboriau-Routhiau V, Bambou JC, et al. (2008) Dissecting the genetic components of adaptation of *Escherichia coli* to the mouse gut. PLoS Genet 4: e2. doi:10.1371/journal.pgen.0040002.

36. Carroll SB (2008) Evo-devo and an expanding evolutionary synthesis: a genetic theory of morphological evolution. Cell 134: 25–36. doi:10.1016/j.cell.2008.06.030.

37. Hoekstra HE, Coyne JA (2007) The locus of evolution: evo devo and the genetics of adaptation. Evolution 61: 995–1016. doi:10.1111/j.1558-5646.2007.00105.x.

38. Hottes AK, Freddolino PL, Khare A, Donnell ZN, Liu JC, et al. (2013) Bacterial adaptation through loss of function. PLoS Genet 9. Available: http://www.ncbi.nlm.nih.gov/pmc/articles/PMC3708842.

39. Maharjan RP, Ferenci T, Reeves PR, Li Y, Liu B, et al. (2012) The multiplicity of divergence mechanisms in a single evolving population. Genome Biol 13: R41. doi:10.1186/gb-2012-13-6-r41.

40. Mason TG, Richardson G (1982) Observations on the *in vivo* and *in vitro* competition between strains of *Escherichia coli* isolated from the human gut. Journal of Applied Microbiology 53: 19–27. doi:10.1111/j.1365-2672.1982.tb04730.x.

41. Gaffé J, McKenzie C, Maharjan RP, Coursange E, Ferenci T, et al. (2011) Insertion sequence-driven evolution of *Escherichia coli* in chemostats. J Mol Evol 72: 398–412. doi:10.1007/s00239-011-9439-2.

42. Rainey, PB, Buckling, A, Kassen, R, Travisano, M (2000) The emergence and maintenance of diversity: insights from experimental bacterial populations. Trends Ecol Evol 15: 243–247.

43. T. Conway, K. A. Krogfelt, P. S. Cohen (2004) The life of commensal *Escherichia coli* in the mammalian intestine. In: R. Curtiss III, editor. EcoSal - *Escherichia coli* and *Salmonella*: cellular and molecular biology. Washington, DC: ASM Press, Vol. Chapter 8.3.1.2. Available: doi: 10.1128/ecosal.8.3.1.2.

44. Miller JH (1996) Spontaneous mutators in bacteria: insights into pathways of mutagenesis and repair. Annu Rev Microbiol 50: 625–643. doi:10.1146/annurev.micro.50.1.625.

45. Giraud A, Radman M, Matic I, Taddei F (2001) The rise and fall of mutator bacteria. Curr Opin Microbiol 4: 582–585.

46. Giraud A, Matic I, Tenaillon O, Clara A, Radman M, et al. (2001) Costs and benefits of high mutation rates: adaptive evolution of bacteria in the mouse gut. Science 291: 2606–2608. doi:10.1126/science.1056421 291/5513/2606.

47. Maharjan RP, Liu B, Li Y, Reeves PR, Wang L, et al. (2013) Mutation accumulation and fitness in mutator subpopulations of *Escherichia coli*. Biol Lett 9: 20120961. doi:10.1098/rsbl.2012.0961.

48. Sniegowski PD, Gerrish PJ, Lenski RE (1997) Evolution of high mutation rates in experimental populations of *E. coli*. Nature 387: 703–705. doi:10.1038/42701.

49. Wylie CS, Trout AD, Kessler DA, Levine H (2010) Optimal strategy for competence differentiation in bacteria. PLoS Genet 6. doi:10.1371/journal.pgen.1001108.

50. Datsenko KA, Wanner BL (2000) One-step inactivation of chromosomal genes in *Escherichia coli* K-12 using PCR products. Proc Natl Acad Sci USA 97: 6640–6645. doi:10.1073/pnas.120163297.

51. Rang CU, Licht TR, Midtvedt T, Conway PL, Chao L, et al. (1999) Estimation of growth rates of *Escherichia coli* BJ4 in streptomycin-treated and previously germfree mice by *in situ* rRNA hybridization. Clin Diagn Lab Immunol 6: 434–436.

52. Leatham-Jensen MP, Frimodt-Møller J, Adediran J, Mokszycki ME, Banner ME, et al. (2012) The streptomycin-treated mouse intestine selects *Escherichia coli envZ* missense mutants that interact with dense and diverse intestinal microbiota. Infect Immun 80: 1716–1727. doi:10.1128/IAI.06193-11.

53. Wilson K (2001) Preparation of genomic DNA from bacteria. Curr Protoc Mol Biol Chapter 2: Unit 2.4. doi:10.1002/0471142727.mb0204s56.

54. Barrick JE, Yu DS, Yoon SH, Jeong H, Oh TK, et al. (2009) Genome evolution and adaptation in a long-term experiment with *Escherichia coli*. Nature 461: 1243–1247. doi:nature08480.

55. Ning Z, Cox AJ, Mullikin JC (2001) SSAHA: a fast search method for large DNA databases. Genome Res 11: 1725–1729. doi:10.1101/gr.194201.

56. Robinson JT, Thorvaldsdóttir H, Winckler W, Guttman M, Lander ES, et al. (2011) Integrative genomics viewer. Nat Biotechnol 29: 24–26. doi:10.1038/nbt.1754.

57. Blattner FR, Plunkett G, Bloch CA, Perna NT, Burland V, et al. (1997) The complete genome sequence of *Escherichia coli* K-12. Science 277: 1453–1462.

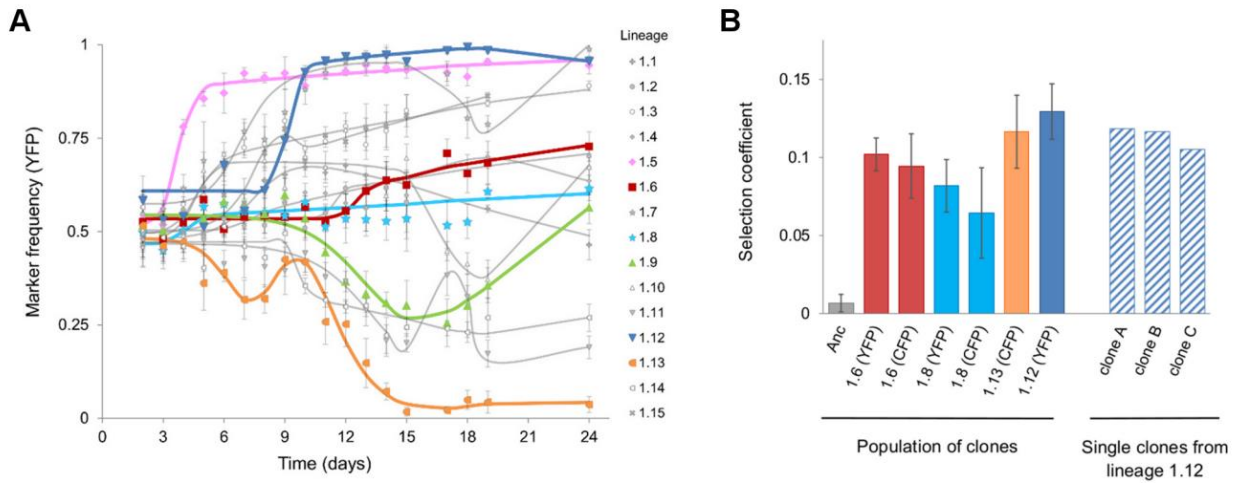

**Figure 1. Evidence for rapid adaptation and CI *in vivo*.**

**A)** Dynamics of marker frequency (with 95% confidence intervals) during the adaptation of *E. coli* to the gut upon an initial colonization (populations 1.1 to 1.15). The predictions of the simplest model of Darwinian selection [22], for each set of data points are shown as lines. The lines correspond to the model that assumes multiple beneficial mutations ($i$=1,2,..,5) can occur in a given clone at a given time ($Tb_i$), and these clones have a given fitness ($W_i$). $Tb_i$ and $W_i$ are fitted by maximum likelihood and the best model, in terms of number of mutations, is chosen according to Akaike criteria. Representative examples of trajectories for the classical signature of a selective sweep (populations 1.5, 1.12 and 1.13) and for the maintenance of neutral diversity under with intense CI (populations 1.6, 1.8 and 1.9) are shown in colours.

**B)** Direct evidence for adaptation and CI: the bars represent the mean selection coefficient (± 2 s.e.m, n=3) from an *in vivo* competitive fitness assay between evolved and ancestral clones. The first bar shows the neutrality of the fluorescent marker. The following six bars represent the results from competitions of a mixture of thirty clones (with a given fluorescent marker, indicated below the bar) isolated from the respective population (indicated below each bar). Both CFP and YFP mixtures of clones from lineages 1.6 and 1.8 show similar levels of adaptation, consistent with intense CI maintaining a high frequency of both neutral markers. The dashed bars show the results of competitions (n=2) of single clones isolated from lineage 1.12.

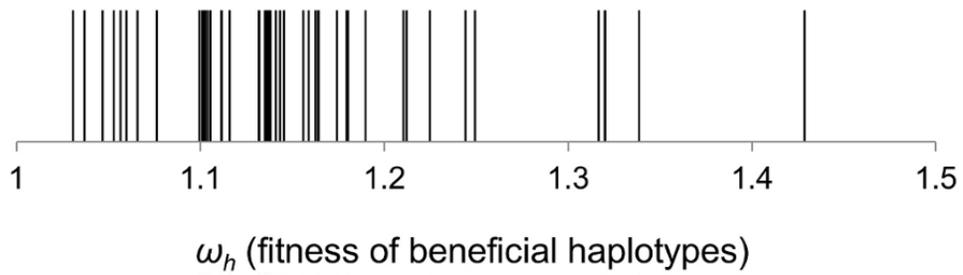

**Figure 2. Distribution of fitness effects of beneficial haplotypes that contributed to adaptation.**

The fitness of beneficial haplotypes ($\omega_h$) was estimated under a theoretical model which assumes the minimum number of beneficial mutations required to explain the marker dynamics.

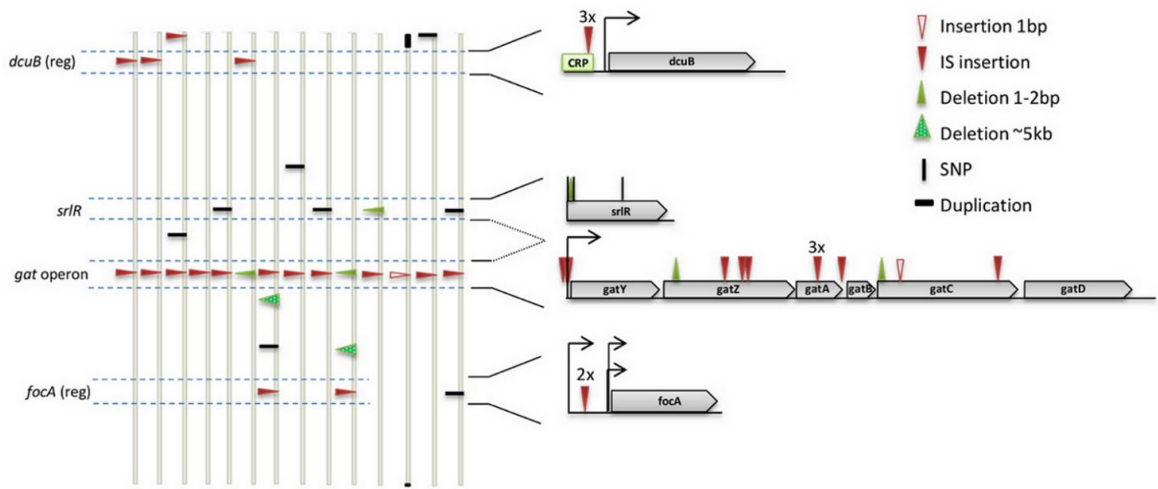

**Figure 3. The genetic basis of adaptive mutations and the level of parallelism between populations.**

Identified mutations in clones isolated from populations 1.1 to 1.14 (evolved *in vivo* for 24 days), represented along the *E. coli* chromosome. For simplicity, the genomes are represented linearly and vertically drawn. The type and position of mutations are shown by triangles for insertions and deletions, small vertical bars denote single nucleotide polymorphisms (SNPs), and one duplication in clone number 1.12 is depicted as a horizontal bar. See the symbol legend for other events. The genes *dcuB*, *srlR* and *focA* and one operon (*gat*) are highlighted. These represent regions of parallel mutation in at least two genomes. The genomic context of these mutations is represented on the right. (reg) after the gene name, means that the regulatory region, rather than the coding region, was affected. Numbers above marked mutations represent the number of times a particular mutation was detected at the same position.

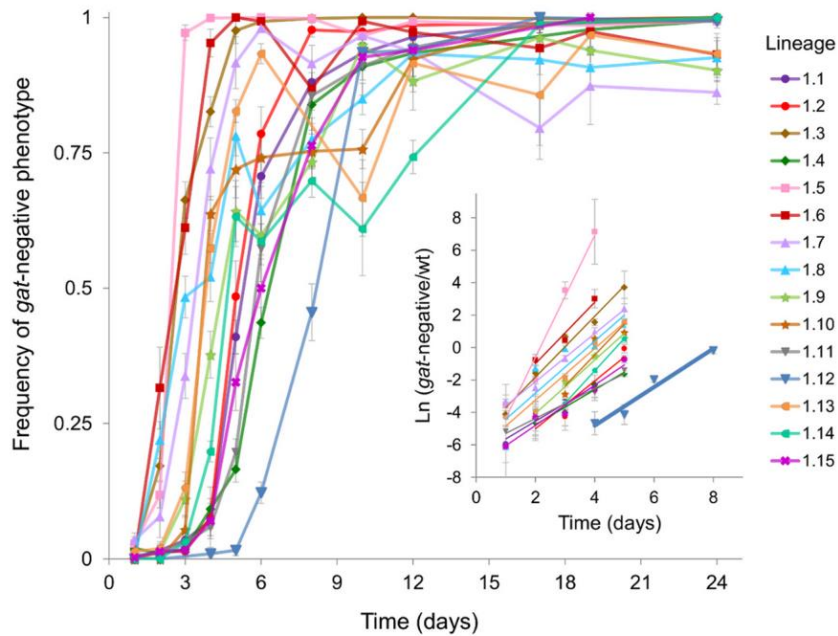

**Figure 4.** Emergence and spread of beneficial mutations in the *gat* operon. Dynamics of frequency change of the *gat*-negative phenotype over time are shown for all populations (1.1 to 1.15). Inset: The natural logarithm of the ratio of *gat*-negative individuals to wild type over the first 5 days of adaptation is shown as dots. Each group of points was fitted to a linear regression (represented as lines). Highlighted in bold is the population 1.12 for which the slope corresponds to an estimate of the selection coefficient of 0.075±0.01 (per generation).

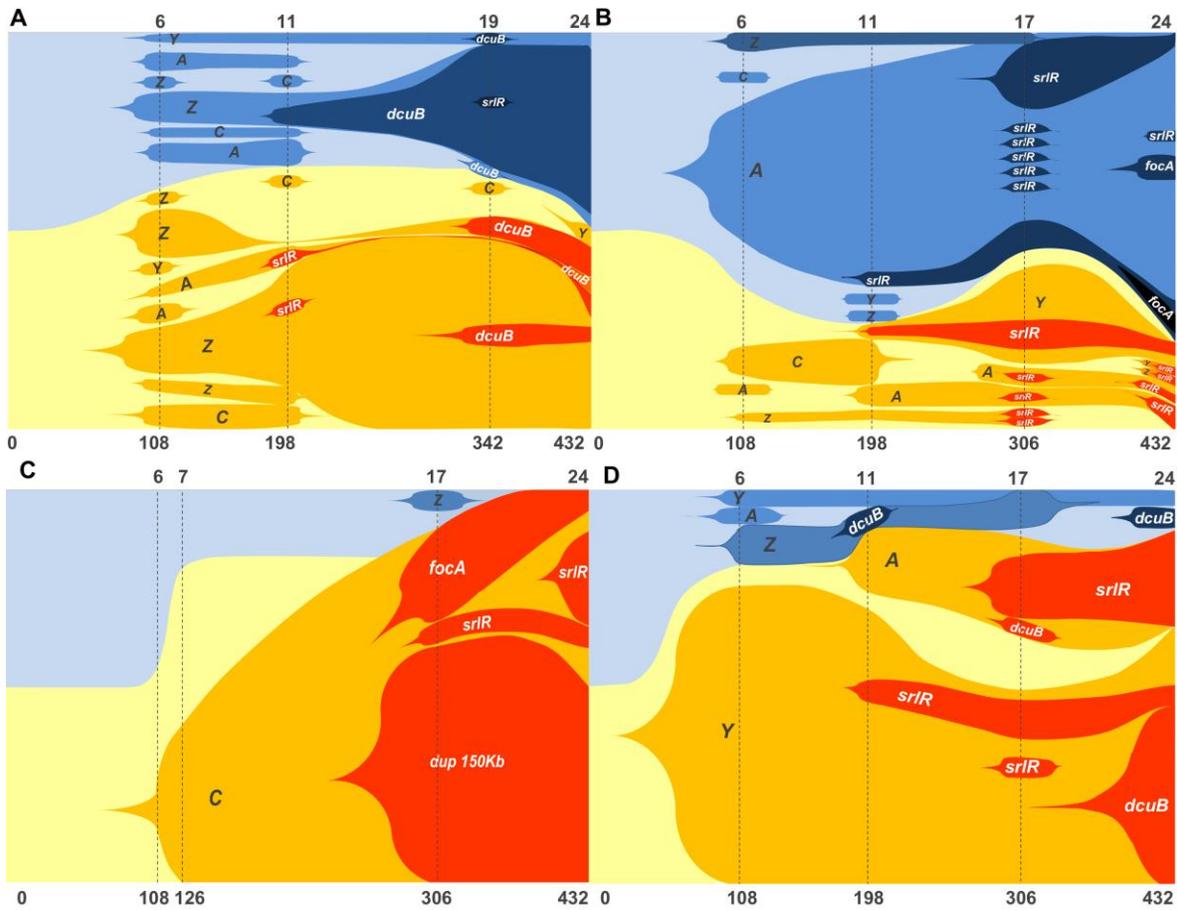

**Figure 5.** Graphic representation of the frequencies of newly generated haplotypes along 24 days (corresponding to 432 generations) of evolution inside the mouse gut (see Tables S3 to S6 for numeric data). Shaded areas are proportional to the relative abundance of each haplotype. Yellow and blue shaded areas represent the two sub-populations of bacteria labeled either with *cfp* or *yfp* alleles. The ancestry relations between haplotypes can be inferred by the accumulation of new mutations in a previously existent genotype. Dash lines mark the time points in days (upper axis) or generations (lower axis) where the sampling took place. For the top two populations 1.1 (A) and 1.11 (B), 40 clones were sampled in each time point. For the bottom two populations 1.12 (C) and 1.5 (D) 20 clones were sampled in each time point.

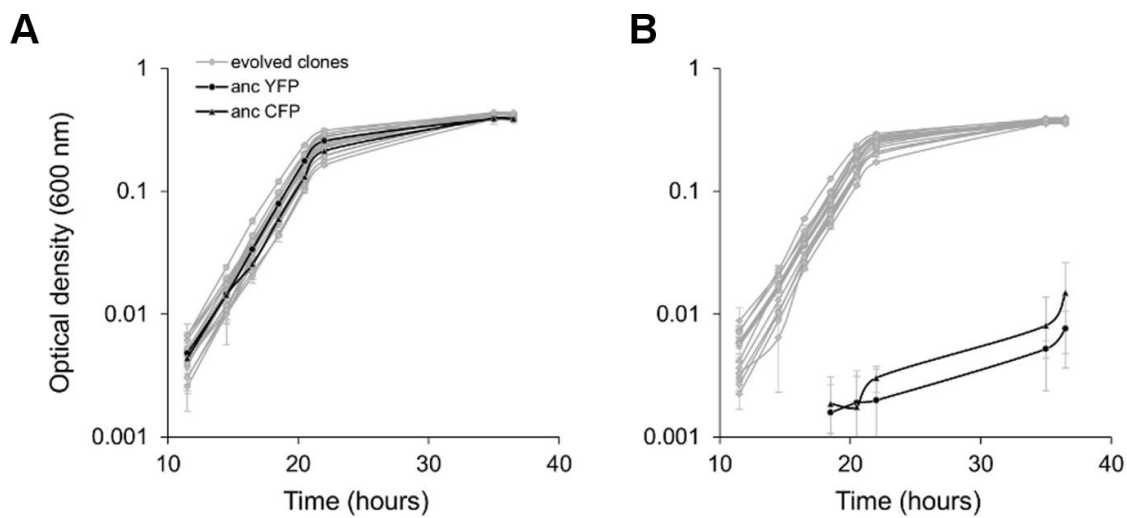

**Figure S1. Comparison of growth curves of ancestral strain and evolved clones.**

Growth curves of ancestral (black) and evolved clones (grey) in MM with glycerol (**A**) and MM with glycerol and galactitol (**B**). Error bars represent the standard error of the mean of three independent measurements.

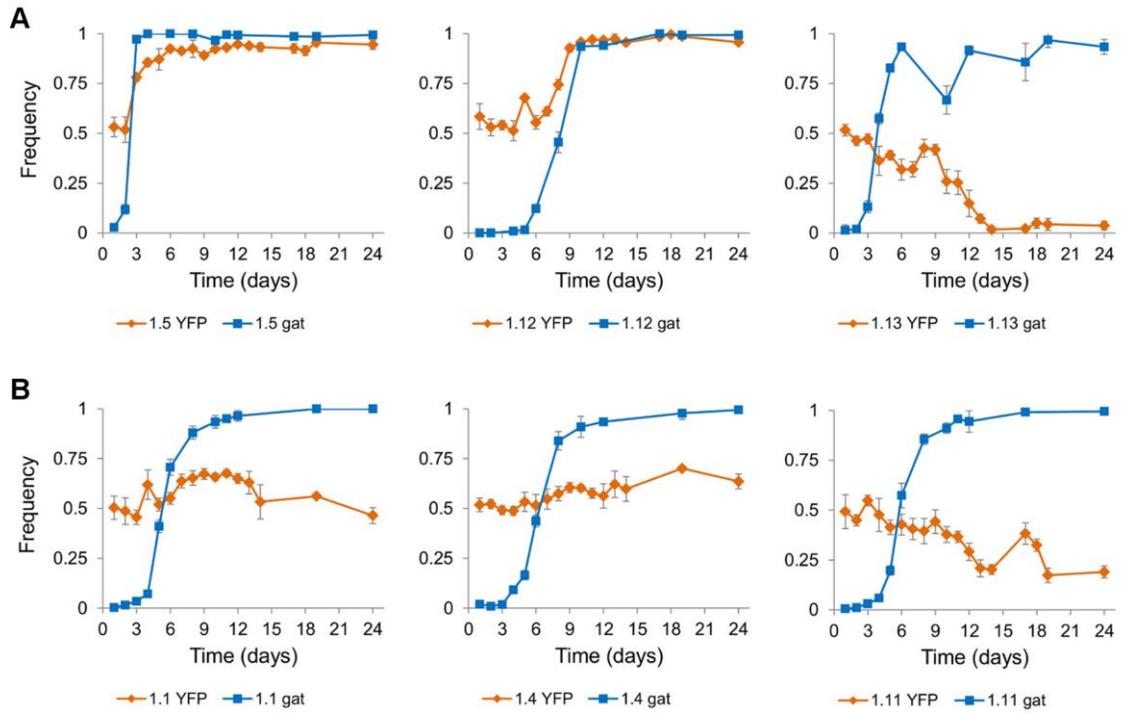

**Figure S2. Emergence and spread of beneficial mutations in the *gat* operon.**

Dynamics of frequency change of the *gat*-negative phenotype (blue squares) and of the neutral fluorescent marker (orange diamonds) are shown for representative examples of populations where increase in frequency and eventual fixation of the *gat*-negative phenotype was accompanied by strong divergence of the fluorescent marker (populations 1.5, 1.12 and 1.13) or not (1.1, 1.4 and 1.11).

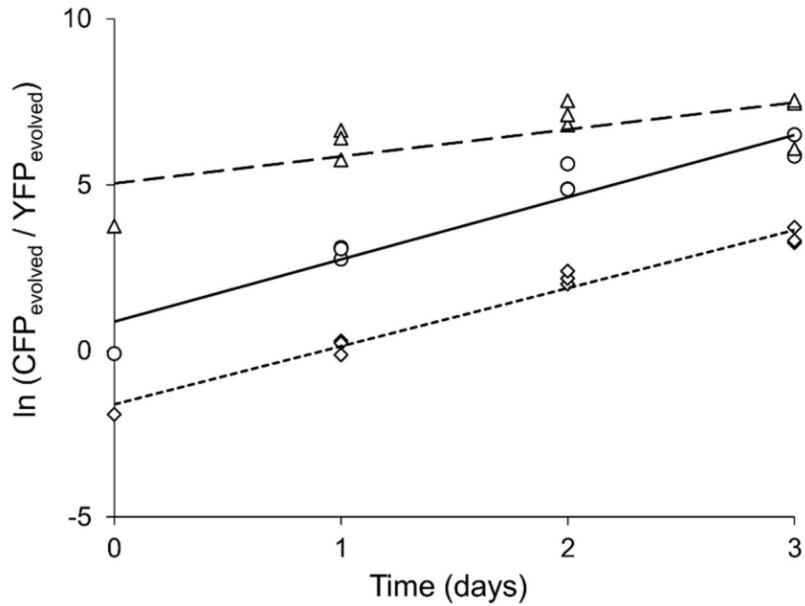

**Figure S3. Test for negative frequency-dependent selection of evolved clones.**

30 CFP and 30 YFP clones isolated from population 1.13 after 24 days of colonization were competed *in vivo* at initial ratios CFP:YFP of 1:1 (open circles), 1:10 (open diamonds) and 100:1 (open triangles) for 3 days (corresponding to approximately 54 generations). Three independent competition experiments were performed for each ratio. The selective advantages per generation of the CFP in relation to the YFP population were calculated for each ratio of CFP over YFP. These are based on the slopes of the linear regression of ln(CFP/YFP) along time. The slopes (±2 s.e.m) are: 0.11 (±0.04), $R^2$=0.95 for 1:1 (dotted line); 0.10 (±0.02), $R^2$=0.99 for 1:10 (solid line) and 0.06 (±0.05), $R^2$=0.77 for 100:1(dashed line). A selective advantage was found for all the cases tested irrespective of their initial frequency, indicating that negative frequency-dependent selection is not the major process underling adaptation in the mouse gut.

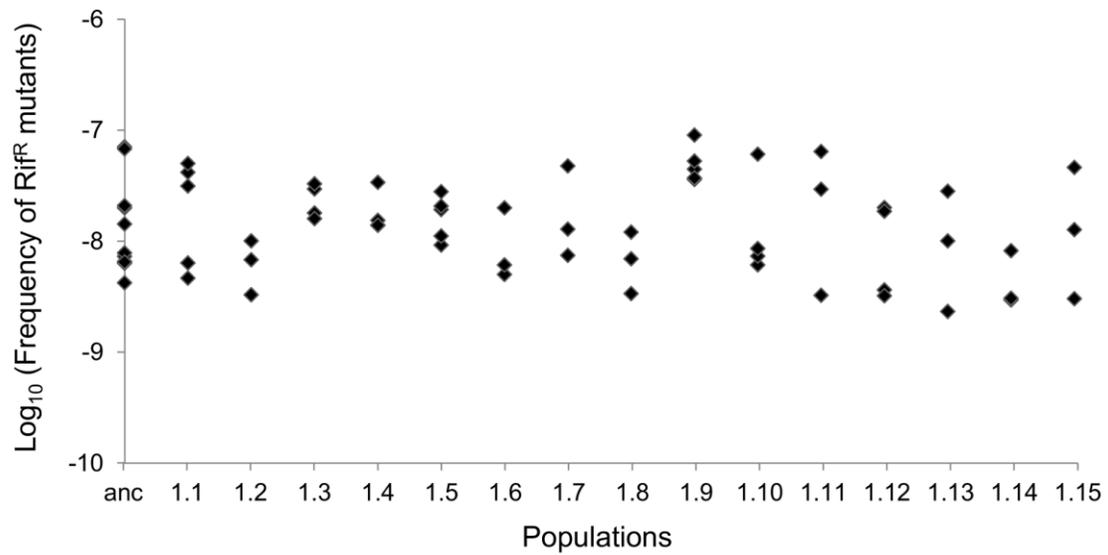

**Figure S4. Fluctuation tests of evolved clones to test for the emergence of mutator alleles during adaptation to the gut.**

Each symbol represents an independent measurement of the frequency of spontaneous mutants resistant to rifampicin. Measurements of the mutation frequency were performed for each of the 15 mice evolved populations (1.1 to 1.15) as well as for the respective ancestors. No significant difference was found between the estimations of the mutation frequencies for the ancestors and evolved populations (Mann-Whitney test with Bonferroni correction, $P>0.05$).

**Table S1. Mutations identified in the genomes of the ancestral strain.** Mutations were identified in the genome of the ancestor 0YFP by comparison with the reference genome[45]. Mutations in intergenic regions have the two flanking genes listed (e.g., fdrA/ylbF). Genes within brackets mean that the mutation happened within the gene. SNPs are represented by an arrow between the ancestral and the evolved nucleotide. Whenever a SNP gives rise to a non-synonymous mutation the amino acid replacement is also indicated. The symbol Δ means a deletion event and a + symbol represents an insertion of the nucleotide that follows the symbol. For intergenic mutations, the numbers in the annotation row represent nucleotides relative to each of the neighboring genes, where + indicates the distance downstream of the stop codon of a gene and - indicates the distance upstream of the gene, that is relative to the start codon. The mutations present in the 0YFP but not in the 0CFP, are underlined.

| Clone | Genome Position | Gene | Mutation | Annotation |
|---|---|---|---|---|
| **0YFP** | 547694 | *fdrA/ylbF* | A→G | intergenic (+123/-1156) |
| | 547835 | *fdrA/ylbF* | +G | intergenic (+264/-1015) |
| | 1395405 | *[ynaJ]–[ttcA]* | Δ13,756 bp | multigenic |
| | 1976527 | *insB–insA* | Δ776 bp | |
| | <u>2369558</u> | <u>*arnT*</u> | <u>4 bp x 2</u> | <u>duplication</u> |
| | 3422257 | *rrlD* | A→C | noncoding |
| | 3422258 | *rrlD* | T→A | noncoding |
| | 3422259 | *rrlD* | C→T | noncoding |
| | <u>3434719</u> | <u>*trkA*</u> | <u>G→A</u> | <u>E60E GAG→GAA</u> |
| | 3472447 | *rpsL* | T→C | K43R AAA→AGA |
| | 3844290 | *uhpT* | A→C | F301V TTT→GTT |
| | 3957957 | *ppiC/rep* | C→T | intergenic (-121/-743) |
| | <u>4095684</u> | <u>*rhaB/rhaS*</u> | <u>T→C</u> | <u>intergenic (-213/-75)</u> |
| | 360473 | *lacA–lacI* | Δ6264 bp | multigenic |
| | 788169 | *[galK]* | Δ1034 bp | |
| | 4294082 | *RIP321* | Δ338 bp | |

**Table S2. Mutations identified in the genomes of the evolved clones.** In the list of mutations, the initials IS denote the abbreviation of insertion sequence element at the indicated position. The asterisk means that the corresponding SNP originated a STOP codon. For further details see Table S1 legend. The last column shows the number of mutations segregating in the lineage from which the sequenced clone was isolated, as inferred from the theoretical model.

| Clone | Genome Position | Gene | Mutation | Annotation | Inferred mutations |
|---|---|---|---|---|---|
| **1CFP** | 2173589 | *gatZ* | IS Ins | coding (755/1263) | 2 |
|  | 4346888 | *dcuB/dcuR* | IS Ins | intergenic (-121/+450) |  |
| **2YFP** | 2173759 | *gatZ* | IS Ins | coding (585/1263) | 2 |
|  | 4346888 | *dcuB/dcuR* | IS Ins | intergenic (-121/+450) |  |
| **3YFP** | 2560011 | *yffN* | G→A | C122Y TGC→TAC | 2 |
|  | 2175242 | *gatY/fbaB* | IS Ins | intergenic (-16/+292) |  |
|  | 4601260 | *yjjP/yjjQ* | IS Ins | intergenic (-379/-244) |  |
| **4YFP** | 2173531 | *gatZ* | IS1 Ins | coding (813/1263) | 3 |
| **5YFP** | 2827493 | *srlR* | G→C | G142A GGC→GCC | 2 |
|  | 2172869 | *gatA* | IS Ins | coding (203/453) |  |
| **6YFP** | 2172262 | *gatC* | del 1 bp | coding (39/1356) | 2 |
|  | 4346888 | *dcuB/dcuR* | IS Ins | intergenic (-121/+450) |  |
| **7YFP** | 1420379 | *ydaV* | C→A | L125M CTG→ATG | 4 |
|  | 1902231 | *[manZ]–[kdgR]* | Δ5,451 bp |  |  |
|  | 2175298 | *gatY/fbaB* | IS Ins | intergenic (-72/+236) |  |
|  | 953904 | *focA/ycaO* | IS Ins | intergenic (-212/+194) |  |
| **8YFP** | 3268729 | *garK* | A→G | F355S TTC→TCC | 2 |
|  | 2172869 | *gatA* | IS Ins | coding (203/453) |  |
| **9YFP** | 2827073 | *srlR* | A→T | K2I AAA→ATA | 2 |
|  | 2172636 | *gatA* | IS Ins | coding (436/453) |  |
| **10CFP** | 2174223 | *gatZ* | Δ2 bp | coding (120-121/1263) | 4 |
|  | 1388754 | *[ycjY-ynaI]* | Δ5315 bp | large deletion |  |
|  | 953904 | *focA/ycaO* | IS5 Ins | intergenic (-212/+194) |  |
| **11CFP** | 2827095 | *srlR* | Δ1 bp | coding (27/774) | 4 |
|  | 2172869 | *gatA* | IS Ins | coding (203/453) |  |
| **12YFP** | 2172079 | *gatC* | +C | coding (222/1356) | 3 |
|  | 4500113 | *[insG-yaaI]* | 2x 151716 bp | large duplication |  |
| **13CFP** | 4637714 | *arcA* | G→T | R206S CGC→AGC | 2 |
|  | 2171153 | *gatC* | IS Ins | coding (1148/1356) |  |
| **14CFP** | 943941 | *dmsC* | G→A | W229* TGG→TAG | 2 |
|  | 2175263 | *gatY/fbaB* | IS Ins | intergenic (-37/+271) |  |
|  | 2827117 | *srlR* | C→T | Q17* CAG→TAG |  |

**Table S3**. Frequencies of newly generated haplotypes along 432 generations of evolution of population 1.1 inside the mouse gut. Yellow and blue shading identify haplotypes belonging to the two sub-populations of bacteria labeled either with *cfp* or *yfp* alleles. The ancestral haplotype is, by definition, devoid of mutations. The genome position for a given mutation is only indicated for mutations that are not IS insertions. Mutations in intergenic regions have the two flanking genes listed (e.g., *dcuB/dcuR*). Whenever a SNP gives rise to a non-synonymous mutation the amino acid replacement is indicated. Synonymous mutations are represented as Syn. The symbol Δ means a deletion event and a + symbol represents an insertion of the nucleotide that follows the symbol. The asterisk means that the corresponding SNP originated a STOP codon. IS insertions at given position are indicated as IS Ins. The exact location of IS insertions is not indicated since it was determined (see Material and Methods), however the detection strategy allowed us to distinguish between different insertions in the same gene. Therefore every haplotype indicated in Table S3 was confirmed to be unique. See Figure 5 for a graphical representation of the data in this table.

| Genome Position | Gene | Mutation | Haplotype frequencies | | | | |
|---|---|---|---|---|---|---|---|
| | | | 0 gen | 108 gen | 198 gen | 342 gen | 432 gen |
| | | | 0.5 | 0.17 | 0.10 | | |
| | *dcuB/dcuR* | IS Ins | | | | 0.03 | |
| | *gatZ* | IS Ins | | 0.07 | 0.05 | | 0.03 |
| | *gatZ* *dcuB/dcuR* | IS Ins IS Ins | | | 0.03 | 0.30 | 0.42 |
| 2827852 | *gatZ* *dcuB/dcuR* *srlR* | IS Ins IS Ins Δ 1bp | | | | 0.03 | |
| 2173887 | *gatZ* | Δ 5bp | | 0.02 | | | |
| 2171894 | *gatC* | Δ 1bp | | | 0.03 | | |
| | *gatY* | IS Ins | | 0.02 | 0.03 | | 0.06 |
| | *gatY* *dcuB/dcuR* | IS Ins IS Ins | | | | 0.03 | |
| | *gatC* | IS Ins | | 0.02 | 0.03 | | |
| | *gatA* | IS Ins | | 0.05 | 0.08 | | |
| | *gatA* | IS Ins | | 0.05 | 0.03 | | |
| | | | 0.5 | 0.15 | 0.08 | 0.03 | |
| | *gatZ* | IS Ins | | 0.12 | 0.38 | 0.46 | 0.28 |
| 2827549 | *gatZ* *srlR* | IS Ins FE161QK | | | 0.03 | | |
| | *gatZ* *dcuB/dcuR* | IS Ins IS Ins | | | | 0.05 | 0.03 |
| | *gatY* | IS Ins | | 0.02 | | | 0.06 |
| | *gatC* | IS Ins | | 0.05 | 0.05 | | |
| 2171454 | *gatC* | Q155* | | | 0.03 | | |
| 2172118 | *gatC* | +C | | | | 0.03 | |
| 2174029 | *gatZ* | ∆1bp | | 0.02 | | | |
| 2173899 | *gatZ* | Δ 5bp | | 0.02 | 0.05 | | |
| 2173900 | *gatZ* | Δ 5bp | | 0.12 | | | |
| 2173900 | *gatZ* | Δ 5bp | | | | 0.05 | 0.08 |
| | *dcuB/dcuR* | IS Ins | | | | | |
| | *gatA* | IS Ins | | 0.05 | | | |
| | *gatA* | IS Ins | | 0.02 | 0.03 | | |
| | *gatA* | IS Ins | | | | | 0.03 |
| | *dcuB/dcuR* | IS Ins | | | | | |
| | *gatA* | IS Ins | | | 0.03 | | 0.03 |
| 2827529 | *srlR* | G154E | | | | | |

**Table S4.** Frequencies of newly generated haplotypes along 432 generations of evolution of population 1.5 inside the mouse gut. See Table S3 for further details.

| Genome Position | Gene | Mutation | Haplotype frequencies | | | | |
|---|---|---|---|---|---|---|---|
| | | | 0 gen | 108 gen | 198 gen | 306 gen | 432 gen |
| | | | 0.5 | | | 0.05 | |
| | *dcuB/dcuR* | IS Ins | | | | | 0.05 |
| | *gatA* | IS Ins | | 0.05 | | | |
| | *gatA* | IS Ins | | 0.10 | | 0.10 | |
| | *gatA* *dcuB/dcuR* | IS Ins IS Ins | | | 0.05 | | |
| | *gatY/fbaB* | IS Ins | | 0.05 | 0.05 | | 0.05 |
| | | | 0.5 | 0.05 | 0.05 | 0.10 | |
| | *gatY* | IS Ins | | 0.75 | 0.65 | 0.40 | 0.15 |
| 2827489 | *gatY* *srlR* | IS Ins P141S | | | 0.05 | 0.05 | |
| 2827490 | *gatY* *srlR* | IS Ins P141L | | | | 0.05 | |
| 2827493 | *gatY* *srlR* | IS Ins G142E | | | | | 0.05 |
| | *gatY* *dcuB/dcuR* | IS Ins IS Ins | | | | | 0.45 |
| | *gatA* | IS Ins | | | 0.15 | 0.05 | |
| 2827493 | *gatA* *srlR* | IS Ins G142A | | | | 0.15 | 0.25 |
| | *gatA* *dcuB/dcuR* | IS Ins IS Ins | | | | 0.05 | |

**Table S5** – Frequencies of newly generated haplotypes along 432 generations of evolution of population 1.11 inside the mouse gut. See Table S3 for further details.

| Genome Position | Gene | Mutation | Haplotype frequencies | | | | |
|---|---|---|---|---|---|---|---|
| | | | 0 gen | 108 gen | 198 gen | 306 gen | 432 gen |
| | | | 0.5 | 0.23 | 0.08 | | |
| | gatA | IS Ins | | 0.33 | 0.50 | 0.15 | 0.57 |
| 2827492 | gatA srlR | IS Ins G142S | | | 0.03 | 0.08 | 0.03 |
| 2827493 | gatA srlR | IS Ins G142D | | | | 0.03 | |
| 2827728 | gatA srlR | IS Ins del1bp | | | | 0.03 | |
| 2827177 | gatA srlR | IS Ins del7bp | | | | 0.18 | 0.03 |
| 2827627 | gatA srlR | IS Ins E187K | | | | 0.03 | |
| 2827076 | gatA srlR | IS Ins P3L | | | | 0.03 | |
| 2827726 | gatA srlR | IS Ins ins7bp | | | | 0.03 | |
| 2827764 | gatA srlR | IS Ins Syn | | | | | 0.03 |
| | gatA focA | IS Ins IS Ins | | | | | 0.05 |
| 2827492 | gatA srlR focA | IS Ins G142S IS Ins | | | | | 0.03 |
| 2173900 | gatZ | Δ 5bp | | 0.05 | 0.03 | 0.03 | |
| | gatC | IS Ins | | 0.03 | | | |
| | gatY | IS Ins | | | 0.03 | | |
| | gatZ | IS Ins | | | 0.03 | | |
| | | | 0.5 | 0.23 | 0.06 | 0.05 | 0.05 |
| 2173878 | gatZ | Δ 5bp | | 0.03 | 0.03 | | |
| 2173878 2827278 | gatZ srlR | Δ 5bp del2bp | | | | 0.03 | |
| 2173878 2827492 | gatZ srlR | Δ 5bp G142S | | | | 0.03 | |
| | gatC | IS Ins | | 0.08 | 0.11 | | |

| | | | | | | |
|---|---|---|---|---|---|---|
| | gatA | IS Ins | | | 0.06 | 0.03 | |
| 2827394 | gatA srlR | IS Ins L109P | | | | 0.03 | |
| 2827496 | gatA srlR | IS Ins G143V | | | | | 0.05 |
| | gatA | IS Ins | | 0.03 | | 0.03 | 0.03 |
| 2827664 | gatA srlR | IS Ins A99E | | | | 0.03 | |
| 2827529 | gatA srlR | IS Ins G154E | | | | | 0.03 |
| 2174150 | gatY | del total | | | 0.03 | 0.13 | 0.03 |
| 2174150 2827172 | gatY srlR | del total T35N | | | 0.03 | 0.10 | 0.03 |
| 2827172 | gatY srlR | IS Ins T35N | | | | | 0.03 |
| 2827276 | gatZ srlR | IS Ins Δ 2bp | | | | | 0.03 |

**Table S6.** Frequencies of newly generated haplotypes along 432 generations of evolution of population 1.12 inside the mouse gut. See Table S3 for further details.

| Genome Position | Gene | Mutation | Haplotype frequencies | | | | |
|---|---|---|---|---|---|---|---|
| | | | 0 gen | 108 gen | 126 gen | 306 gen | 432 gen |
| | | | 0.5 | 0.48 | 0.20 | 0.05 | |
| 2172078 | gatC | +C | | | | 0.05 | |
| | | | 0.5 | 0.42 | 0.40 | | |
| | gatC | +C | | 0.09 | 0.40 | | 0.15 |
| | gatC dup | +C | | | | 0.60 | 0.50 |
| | gatC focA | +C IS Ins | | | | 0.25 | 0.05 |
| 2827117 | gatC srlR | +C Q17* | | | | 0.05 | 0.05 |
| 2827234 | gatC srlR | +C L97V | | | | | 0.25 |

**Table S7. Oligonucleotide primers used in this work.**

| Gene | Forward (5' - 3') | Reverse (5' - 3') | Usage |
|---|---|---|---|
| *dcuB* | GGCTGAAGGTGGAAGACGAA | ACATTTCGCGTGTTTCCTGC | Amplification of *dcuB* |
| *focA* | AGCGGATGTTTCGTTGCTTT | TGCTGCACATCAGTCGTTGT | Amplification of *focA* |
| *gatABCD* | TCCCACCGCATCAATATAGCC | CAGTCCGGGGAATTATCAGCA | Amplification of *gatA, gatB, gatC and gatD* |
| *gatZY* | CACCTTTGGCGAGCATCTCA | AAAACACGCGCACTTTGCTA | Amplification of *gatZ and gatY* |
| *gatB* | GATCCACTTTGGCAGTGGTG | CAGTCCGGGGAATTATCAGCA | Amplification of *gatB* |
| *gatY* | GCCACAATCGGCAATCACTT | AAAACACGCGCACTTTGCTA | Amplification of *gatY* |
| *dupl 150kb* | GTTCGTTGCGCATCAGTACG | GCTCTACCCGCAGGTCAAAA | Amplification of the new junction |
| *srlR* | GCATGCGGGTGATTTACAGC | TTCCGGTAAACGGCTTGCTT | Amplification/sequencing of *srlR* |
| *gatC* | ATTAGCCGCCAGTTGGGTG | | sequencing of gatC |
| | TGCCGATAATCAGCCCCATC | | sequencing of gatC |
| | CCAGCCAACATCGACCACAT | | sequencing of gatC |
| *gatZ* | ATATCGCCTCGCGTAAAGCA | | sequencing of gatZ |
| | ACCGTTTCTGGTGCTAACGG | | sequencing of gatZ |